\documentclass[prb,twocolumn,showpacs,amsmath]{revtex4} 

\usepackage{graphicx}
\usepackage{hyperref}

\pacs{71.15.-m, 71.15.Ap, 71.15.Mb, 31.15.E-}

\DeclareMathOperator{\erf}{erf}
\DeclareMathOperator{\erfc}{erfc}
\DeclareMathOperator{\spherical}{Y}
\providecommand{\unit}[1]{\ensuremath{\,\text{#1}}}
\providecommand{\pedex}[1]{\ensuremath{_\text{#1}}}
\providecommand{\apex}[1]{\ensuremath{^\text{#1}}}
\providecommand{\etal}{\emph{et~al.}}
\providecommand{\vect}[1]{\ensuremath{\mathbf{#1}}}
\providecommand{\basisvec}[1]{\ensuremath{\widehat{\vect #1}}}
\providecommand{\myeqref}[1]{Eq.~\eqref{#1}}
\providecommand{\myfigref}[1]{Fig.~\ref{#1}}
\providecommand{\mytabref}[1]{Table~\ref{#1}}
\providecommand{\eunit}{\ensuremath{\text{e}}}
\providecommand{\imag}{\ensuremath{\text{i}}}
\providecommand{\expp}[1]{\ensuremath{\mathop{\eunit^{#1}}}}
\providecommand{\abs}[1]{\ensuremath{\left\lvert#1\right\rvert}}
\providecommand{\diff}[2][]{\ensuremath{\mathop{\text{d}^{#1}{#2}}}}
\providecommand{\conjg}[1]{\ensuremath{{#1}^\ast}}
\providecommand{\mixed}{\ensuremath{M}}
\providecommand{\potential}{\ensuremath{V}}
\providecommand{\hamiltonian}{\ensuremath{\mathcal H}}

\makeatletter
\providecommand{\bra}{\@ifstar\bra@star\bra@nostar}
\providecommand{\bra@nostar}[1]{\ensuremath{\left\langle #1 \right\rvert}}
\providecommand{\bra@star}[1]{\ensuremath{\langle #1 \rvert}}
\providecommand{\ket}{\@ifstar\ket@star\ket@nostar}
\providecommand{\ket@nostar}[1]{\ensuremath{\left\lvert #1 \right\rangle}}
\providecommand{\ket@star}[1]{\ensuremath{\lvert #1 \rangle}}
\providecommand{\braketmat}{\@ifstar\braketmat@star\braketmat@nostar}
\providecommand{\braketmat@nostar}[3]{\ensuremath{\left\langle #1 \left\lvert #2 \right\rvert #3 \right\rangle}}
\providecommand{\braketmat@star}[3]{\ensuremath{\langle #1 \lvert #2 \rvert #3 \rangle}}
\providecommand{\braketvecLeft}[2]{\ensuremath{\left\langle\left. #1 \right\rvert #2 \right\rangle}}

\providecommand{\braketvec}{\@ifstar\braketvec@star\braketvecLeft}
\providecommand{\braketvec@star}[2]{\ensuremath{\langle #1 \vert #2 \rangle}}
\makeatother

\begin{document}

\title{The HSE hybrid functional within the FLAPW method and its application to GdN}
\author{Martin Schlipf}
\email{m.schlipf@fz-juelich.de}
\author{Markus Betzinger}
\author{Christoph Friedrich}
\author{Marjana Le\v{z}ai\'{c}}
\author{Stefan Bl\"ugel}
\affiliation{Peter Gr\"unberg Institut and Institute for Advanced Simulation,
  Forschungszentrum J\"ulich and JARA, 52425 J\"ulich, Germany}

\begin{abstract}
We present an implementation of the Heyd-Scuseria-Ernzerhof (HSE) hybrid functional
within the full-potential linearized augmented-plane-wave (FLAPW) method.
Pivotal to the  HSE functional is the  screened electron-electron interaction,
which we separate into the bare Coulomb interaction and the
remainder, a slowly varying function in real space.
Both terms give rise to exchange potentials, which sum up to the
screened nonlocal exchange potential of HSE.
We evaluate the former with the help of an auxiliary basis,
defined in such a way that the bare Coulomb matrix becomes sparse.
The latter is computed in reciprocal space,
exploiting its fast convergence behavior in reciprocal space.
This approach is general and can be applied to a whole class
of screened hybrid functionals.
We obtain excellent agreement of band gaps and lattice constants for prototypical semiconductors and
insulators with electronic-structure calculations using plane-wave or Gaussian basis sets. We apply the HSE
 hybrid functional to examine the ground-state properties of rocksalt GdN, which have been controversially
discussed in literature. Our results indicate that there is a half-metal to insulator transition occurring between the
theoretically optimized lattice constant at $0\unit{K}$ and the experimental lattice constant at room temperature. Overall,
we attain good agreement with experimental data for band transitions, magnetic moments, and the Curie temperature.
\end{abstract}

\maketitle

\section{Introduction}

Density functional theory (DFT)\cite{hk64,ks65} is a powerful tool for calculating the electronic ground-state properties
 of molecules and solids. The predictive power of numerical DFT calculations relies on the availability of accurate approximations
 for the exchange-correlation (xc) energy $E\pedex{xc}$, which incorporates all complicated many-body effects.
 In many systems, this quantity is described adequately by the local-density approximation (LDA).\cite{bh72,gl76,ca80}
However, in more complex systems, physical properties such as the geometric structure, magnetic properties, and the band
 gap are not well reproduced. One can go beyond the LDA by taking into account the local density gradient, which yields
the generalized gradient approximation (GGA),\cite{pw86,pbe96} upon which many functionals are based. However, despite
their success, the GGA functionals still often fail in describing systems with localized states, which is attributed
to an incomplete cancellation of the self-interaction error in these semi-local functionals.\cite{fa34,pz81}
\par

This deficiency is particularly critical in systems whose electronic properties are largely governed by the
correlated motion of electrons in localized states. The rare-earth chalcogenides are among this class of materials,
having incompletely filled $f$-electron shells. They are insulating, semiconducting, or metallic depending on details
of the valency of the rare-earth element. Gadolinium nitride (GdN) is
widely studied owing to the ferromagnetic order, large magnetic moment of $6.88\unit{$\mu\pedex{B}$}$
 per Gd atom\cite{liss94} and its large magnetoresistive effect,\cite{leuen05}
which makes the material interesting for technological applications.
The mechanism of the ferromagnetic order is still under debate. Various
types are being discussed, such as carrier mediated\cite{ml08,sn10} and superexchange mechanisms.\cite{duanall}
Another point of debate are the electronic properties. It was experimentally demonstrated to be a low carrier
semimetal in single crystals\cite{wk80} and insulating in thin films.\cite{xc96} There are also several recent
reports of thin films of GdN having a degenerately doped semiconducting\cite{gra06,lud09,sca09}
or a metallic ground state\cite{sca09} based on the resistivity data measured at
low temperatures.
Theoretically it is predicted to have a
semiconducting\cite{lam00,gdd05} or a half-metallic character\cite{aer04,duanall,doll08} based on {\it ab
initio} calculations.
\par

Materials with strongly localized
states, such as the $f$ states in GdN, are often treated within the LSDA+$U$ method,\cite{aal97} where the electron
correlation in these states is described with an additional on-site term, that involves the Hubbard
parameter $U$. The disadvantage of this method is that the value of $U$ is not known a priori. Although methods to
estimate the $U$ value from first-principles calculations have been developed,\cite{dbza84,cg05,sfb11} it is usually chosen to
reproduce experimental observations.
However, a specific $U$ value that provides a good description of one quantity is often not suitable to describe
another quantity.\cite{rhk03}
\par

During the last decade, hybrid functionals that combine a fraction of nonlocal Hartree-Fock (HF) exchange with local xc
functionals have been shown to be a viable improvement over LDA and GGA offering a parameter-free description specifically
suited for band gap materials.\cite{mwh01,brothers08} The explicit consideration of
nonlocal HF exchange leads to a partial cancellation of the self-interaction
error, but also makes numerical calculations considerably more demanding
than conventional LDA and GGA calculations. Various hybrid functionals have
been developed. In empirical hybrid functionals, such as the B3LYP functional,\cite{bec2_93} the fraction of HF exchange is
determined by fitting to an experimental data set. In the PBE0 functional\cite{pbe096} the mixing
parameter for the HF exchange is inferred from expanding the integrand of the adiabatic-connection formula of the
exact xc functional.

In periodic systems, the Coulomb interaction between the electrons is effectively screened by polarization effects
in the electron system.
The effective interaction is particularly short-ranged in systems with small or vanishing band gaps.
Therefore, Heyd \etal\cite{hse03} introduced a range-separated hybrid functional, which has the added benefit
to reduce the computational cost within a basis of localized Gaussian functions. Starting
from the PBE0 functional, they partitioned the Fock exchange term into a short- and a long-range part, where the former is
described by a correspondingly screened Fock term and the latter is treated by a local approximation, derived from the
PBE functional.\cite{pbe96} Heyd \etal~showed that  this hybrid functional leads to a reduced computational demand for
localized basis sets compared with the PBE0 functional. Furthermore, it even yields results which are often
in better agreement with experiment.\cite{brothers08}\par

The Heyd-Scuseria-Ernzerhof (HSE) hybrid functional has been implemented within Gaussian \cite{hse03} and
plane-wave\cite{phmk05,paier06} basis sets. In this work, we present an implementation within
the full-potential linearized augmented-plane-wave (FLAPW) approach
as implemented in the \texttt{Fleur} code,\cite{fleur}
which provides a highly accurate all-electron basis\cite{wkwf81,wwf82,jf84} for a large
variety of materials, including open systems with low symmetry, $d$- and $f$-electron systems, as well as oxides
and nitrides.
An implementation of the PBE0 functional limited to certain localized states and on-site interactions
was given by Tran \etal.\cite{tbsn06} Betzinger \etal\cite{bfb10} described an efficient way to calculate
the full nonlocal exchange potential for the PBE0 functional without these
restrictions.
Very recently, Tran and Blaha\cite{tb11} reported an implementation of hybrid
functionals whose nonlocal exchange integrals are
evaluated using the pseudocharge method of Weinert.\cite{wei81}
However, this approach mathematically restricts the electron-electron interaction to potentials
that are solutions of Laplace-type equations, whose radial solutions can be
expressed as analytically or numerically known regular and irregular solutions
and spherical harmonics,
such as the bare Coulomb and the screened Yukawa
potential. The error function used in the HSE functional does not have this
property. In our implementation, there is no such restriction.
Our approach is very general. In fact, any kind of interaction potential can be implemented for the
nonlocal exchange potential by changing only a
single line of code. The only requirement is that it differs from the bare
Coulomb potential by a function that possesses a fast Fourier expansion,
a condition that is fulfilled by all physical screened potentials, including
the error function used in the HSE functional.

\par
Our numerical approach extends the implementation of
Ref.~\onlinecite{bfb10}, which is based on an auxiliary basis that is designed to represent products of
wave functions. This so-called mixed product basis (MPB) is constructed directly from products of LAPW basis functions and retains the
full accuracy of the all-electron description. Several techniques were introduced to accelerate the evaluation of the
computationally expensive nonlocal exchange term. Spatial and time-reversal symmetries are exploited to restrict the
Brillouin-zone (BZ) summations to irreducible sets of $\vect k$-points. The nonlocal potential is calculated in the basis of
single-particle eigenstates, which allows to truncate the matrix at a certain number of bands. The divergence of the
Coulomb potential in the BZ center is treated analytically instead of using dense $\vect k$-point sets around
$\vect k=\vect 0$. A nested density convergence scheme greatly reduces the number of iterations in the self-consistent
field cycle. Finally, by a suitable transformation of the MPB the Coulomb matrix becomes sparse, which speeds up
the matrix-vector operations considerably.
\par
This transformation relies on the analytic properties of the bare Coulomb potential. Any other potential, in particular
the screened Coulomb interaction, will not lead to a sparse matrix representation, though. Furthermore, in contrast to
Gaussian or plane-wave basis sets, a direct evaluation of the screened Coulomb matrix, in the same way as for the bare
Coulomb matrix,\cite{fsb09} is cumbersome in the MPB. Therefore, we incorporate the screening,
after calculating the bare nonlocal exchange potential, in a separate step, which produces hardly any overhead.
In this way, the simple analytic properties of the bare Coulomb potential as well as
the sparsity of the Coulomb matrix are retained and can be taken advantage of.\par
We validate our implementation by comparing results for prototypical semiconductors and insulators with
results from the literature. Then, we calculate the ground-state properties and the band structure of GdN.
The band gap of GdN is controversially discussed in literature.
Results from LSDA+$U$ calculations are inconclusive.
While the linearized muffin-tin orbital (LMTO) approach yields a narrow-gap semiconductor as ground state,\cite{lam00,ll06,llcs07,ml08} GdN
exhibits a transition from a half-metallic to a semiconducting ground state under strain within the FLAPW
method.\cite{duanall} Two different solutions close in energy were obtained in an investigation using the hybrid
functional B3LYP.\cite{doll08} Both solutions were half-metallic, one in the majority spin channel, the other one in the
minority spin channel. Our results show a transition from a half-metallic ground state, which is similar to the
energetically slightly less favorable solution of Ref.~\onlinecite{doll08}, to a semiconductor under strain as in the
LSDA+$U$ calculation of Ref.~\onlinecite{duanall}. In their work, a large change of the lattice constant of more than
$10\%$ was necessary to observe this transition. However, our calculations indicate that already small volume changes
($\approx 0.5\%$) are sufficient to observe this transition.\par
The paper is organized as follows. In Sec.~\ref{sec:theory}, we give a brief introduction to the theory of hybrid
functionals. In Sec.~\ref{sec:impl}, we introduce the FLAPW method and describe our implementation of the
HSE functional. In Sec.~\ref{sec:calc}, we first compare results for prototypical semiconductors and insulators
with values from the literature. Then, we present our findings for GdN in Sec.~\ref{sec:gdn}.
 In Sec.~\ref{sec:conclusion}, we draw our conclusions.

\section{Theory}\label{sec:theory}
The construction of hybrid functionals as mixtures of local functionals with a nonlocal exchange term is motivated by
the adiabatic-connection formula,\cite{hj74,gl76,lp77} in which the scaling
of the electron-electron interaction establishes a connection between the
noninteracting Kohn-Sham system with the fully interacting one. In the
weakly interacting limit, the formula becomes identical to the HF exchange
term which prompted Becke\cite{bec2_93,bec96} to introduce a certain fraction $a$ of HF exchange
into the xc functional
\begin{equation} \label{eq:hybrid}
	E\pedex{xc}\apex{HYB} = E\pedex{xc}\apex{L} + a \left( E\pedex{x}\apex{HF} - E\pedex{x}\apex{L} \right)\space,
\end{equation}
where $E\pedex{xc}\apex{L}$ denotes the local xc functional and $E\pedex{x}\apex{L}$ its exchange part. $E\pedex{x}\apex{HF}$ is the HF exchange energy
	\begin{multline} \label{eq:HF_en}
		E\pedex{x}\apex{HF} = - \frac 12 \sum_{\sigma}\sum_{\vect k,\vect q}\apex{BZ}\sum_{n,n'}\apex{occ.}
	                        \iint \diff[3]r \diff[3]{r'} \\
				{\varphi_{n\vect k}^{\sigma\ast}}(\vect r)\varphi_{n'\vect q}^\sigma(\vect r)
				v(\abs{\vect r-\vect r'})
				{\varphi_{n'\vect q}^{\sigma\ast}}(\vect r')\varphi_{n\vect k}^\sigma(\vect r')\space,
	\end{multline}
evaluated with the Kohn-Sham wave functions $\varphi_{n\vect k}^\sigma(\vect r)$ of spin $\sigma$, wave vector $\vect
k$, and band index $n$. The sums over $\vect k$ and $\vect q$ run over the full Brillouin zone (BZ),
 $n$ and $n'$ sum
over all occupied states, and $v(r) = 1/r$ is the bare Coulomb potential. Here and
in the following, atomic units are used unless stated otherwise.
As the wave functions $\varphi_{n\vect k}^\sigma(\vect r)$ are functionals of the effective potential, which in turn is
a functional of the density, $E\pedex{x}\apex{HF}$ is a true functional of the density, too.
\par

Perdew \etal\cite{pbe096} deduced a mixing parameter $a = 0.25$ by assuming a certain shape for the adiabatic-connection
integrand. They proposed to use the Perdew-Burke-Ernzerhof (PBE) functional\cite{pbe96} for the local part.
The resulting functional
\begin{equation} \label{eq:pbe0}
	E\pedex{xc}\apex{PBE0} = E\pedex{xc}\apex{PBE} + a \left( E\pedex{x}\apex{HF} - E\pedex{x}\apex{PBE} \right)
\end{equation}
is nowadays referred to as PBE0.
\par
As the long-range  part of the nonlocal HF term is cumbersome to calculate, Heyd \etal\cite{hse03,jhs09} suggested to
replace it again by a simple local functional. Later, it was demonstrated\cite{brothers08} that this leads to an
improved description of the band gaps of semiconductors. Heyd \etal\cite{hse03} used the error function $\erf(x)$ and
its complement $\erfc(x) = 1 - \erf(x)$ to decompose the Coulomb interaction into a long-range (LR) and a short-range (SR) part
\begin{align} \label{eq:potSep}
	v(r) = \frac{\erf(\omega r)}{r} + \frac{\erfc(\omega r)}{r} = v\apex{LR}(r) + v\apex{SR}(r)\space,
\end{align}
where $\omega$ is an adjustable screening parameter. The HSE hybrid functional is thus given by
\begin{equation} \label{eq:enHSE}
	E\pedex{xc}\apex{HSE}(\omega) = E\pedex{xc}\apex{PBE} + a \left[ E\pedex{x}\apex{HF,SR}(\omega) - E\pedex{x}\apex{PBE,SR}(\omega) \right]\space,
\end{equation}
where $E\pedex{x}\apex{HF,SR}(\omega)$ corresponds to \myeqref{eq:HF_en} with the bare Coulomb potential
 $v(r)$ replaced by $v\apex{SR}(r)$. $E\pedex{x}\apex{PBE,SR}(\omega)$ is the local
 functional for the SR exchange according to the decomposition given in \myeqref{eq:potSep}.
 Its numerical treatment is discussed in
 Refs.~\onlinecite{hse03} and \onlinecite{hs04}.
Based on numerical fits to benchmark data sets of molecules, Heyd \etal\cite{hse03} found an optimized value
 for the screening parameter $\omega = 0.15$.
In this work, we employ the value of $\omega = 0.11$, which was optimized for solids.\cite{kvis06}\par

Hybrid functionals are usually applied within the generalized Kohn-Sham formalism,\cite{seidl96} which is based on a
fictitious system of noninteracting electrons. These particles move subject to a nonlocal potential that is defined
in such a way that the electron density $n(\vect r)$ equals that of the real system. The nonlocal potential contains
a local part that consists of the external potential created by the nuclear charges, the Hartree potential, i.e., the
electrostatic potential produced by the total electron charge density, and a local contribution
that derives from functional differentiation of the local parts of
\myeqref{eq:pbe0} and \eqref{eq:enHSE} for the PBE0 and HSE functionals, respectively.
The implementation of this local part of the xc potential requires only
minor modifications of the DFT code, and we will focus on the nonlocal part in the
following, which derives from the nonlocal exchange energy functional $E\pedex{x}\apex{HF}$.
{Leaving out the scaling factor $a$,} its matrix representation in the basis of Kohn-Sham eigenstates is given by
	\begin{multline} \label{eq:pot_nl}
		V_{\text{x},nn'}^{\sigma,\text{NL}}(\vect k) = - \sum_{\vect q}\apex{BZ}\sum_m\apex{occ.}
		\iint \diff[3]r \diff[3]{r'} \\ \times
		{\varphi_{n\vect k}^{\sigma\ast}}(\vect r){\varphi_{m\vect q}^{\sigma}}(\vect r)
		v(\abs{\vect r-\vect r'}) \varphi_{m\vect q}^{\sigma\ast}(\vect r') \varphi_{n'\vect k}^\sigma(\vect r')\space.
	\end{multline}
For the HSE functional, the bare interaction $v(r)$
would have to be replaced by the screened interaction $v\apex{SR}(r)$. Yet in practice, we first evaluate the
nonlocal potential in the form of Eq.~(\ref{eq:pot_nl}) and correct for the
screening afterwards by subtracting $v\apex{LR}(r)$, as will be explained in the next section.

\section{Implementation} \label{sec:impl}

\subsection{Basis sets}
Our implementation of the HSE functional is based on the all-electron FLAPW method,\cite{wkwf81,wwf82,jf84} in
which space is partitioned into non-overlapping atom-centered muffin-tin (MT) spheres and the remaining interstitial region (IR). The core states,
which are confined to the spheres, are obtained from solving the fully relativistic Dirac equation. For the valence and
conduction states we use a set of basis functions that are defined differently in the two regions of space: plane waves
$\expp{\imag(\vect k+\vect G) \cdot \vect r}$ with $\abs{\vect k+ \vect G} \le G\pedex{max}$ in the IR and linear
combinations of $u^{a\sigma}_{lp}(r) Y_{lm}(\basisvec r)$ in the MT spheres, where
$r$ is measured from the sphere center, $u^{a\sigma}_{lp}(r)$ are numerical functions defined on a radial grid,
$Y_{lm}(\basisvec r)$ are spherical harmonics with angular-momentum quantum
numbers $0 \le l \le l\pedex{max}$ and $\abs{m} \le l$, $a$ labels the atom in the unit cell, and $p$ is an index for
different radial functions. $G\pedex{max}$ and $l\pedex{max}$ are cutoff parameters. The linear combinations are such
that the basis functions and their radial derivatives are continuous at the MT sphere boundaries.\par

As in Ref.~\onlinecite{bfb10}, we evaluate the nonlocal potential, \myeqref{eq:pot_nl}, with the help of an auxiliary basis
$\{M_{\vect qJ}(\vect r)\}$ and its bi-orthogonal set $\{\tilde M_{\vect qJ}(\vect r)\}$, where $\vect q$ is a Bloch vector
and $J$ is used to index these basis functions.
By placing the completeness relation
\begin{equation}
	1 = \sum_{\vect qJ} \ket{M_{\vect qJ}}\bra*{\tilde M_{\vect qJ}} = \sum_{\vect qJ} \ket*{\tilde M_{\vect qJ}}\bra{M_{\vect qJ}}
\end{equation}
at both sides of $v(r)$, the six-dimensional integral becomes a sum over vector-matrix-vector products in the space of the MPB
	\begin{multline} \label{eq:introMixed}
		\potential_{\text{x},nn'}^{\sigma,\text{NL}}(\vect k) =
		-\sum_{m}\apex{occ.}\sum_{\vect q}\apex{BZ} \sum_{IJ}
		\braketvec*{ \varphi^\sigma_{n\vect k} }{
		\varphi^\sigma_{m\vect k - \vect q} \mixed_{\vect q,I} } \\
		\times v_{IJ}(\vect q) \braketvec*{ \mixed_{\vect q,J} \varphi^\sigma_{m\vect k - \vect q}}{ \varphi^\sigma_{n'\vect k} },
	\end{multline}
with the usual bra-ket notation $\braketvec{f}{g} = \int \diff[3]r \conjg{f}(\vect r) g(\vect r)$ and the Coulomb matrix\cite{fsb09}
\begin{equation} \label{eq:coulMat}
	v_{IJ}(\vect q) = \iint \diff[3]r \diff[3]{r'} \tilde{\mixed}_{\vect q,I}^{\ast}(\vect r) v(\vect r,\vect r') \tilde \mixed_{\vect q,J}(\vect r')\space.
\end{equation}
We need to evaluate the latter only once at the beginning of the
self-consistency cycle. We note here again that the screening is accounted
for in a separate step [cf.~Eq.~(\ref{eq:sepHSE})].\par

As \myeqref{eq:introMixed} indicates, the auxiliary basis must be sufficiently complete in the space of wave-function
products. We therefore construct this basis directly from products of LAPW basis functions. In
the interstitial region, the auxiliary basis functions are plane waves with a new cutoff parameter $G'\pedex{max}$ and
 in the MT spheres, the basis consists of numerical functions of the form
$\mixed^a_{LP}(r) \spherical_{LM}(\basisvec r)$ with a cutoff $L\pedex{max}$. This so-called
mixed product basis (MPB) can be systematically converged to represent the products of LAPW wave functions numerically
 exactly. In this
way, the MPB is on the same level of accuracy as the all-electron LAPW basis for the wave functions. It was
shown\cite{bfb10,fbs10} that $G'\pedex{max}$ and $L\pedex{max}$ can be well below their mathematically determined exact limit,
 i.e., $2 G\pedex{max}$ and $2 l\pedex{max}$, and even below $G\pedex{max}$ and $l\pedex{max}$, which makes the MPB a very
efficient basis. Further details about the MPB can be found elsewhere.\cite{bfb10,fsb09,fbs10}

\subsection{Implementation of the HSE functional}

\begin{figure}
	\includegraphics{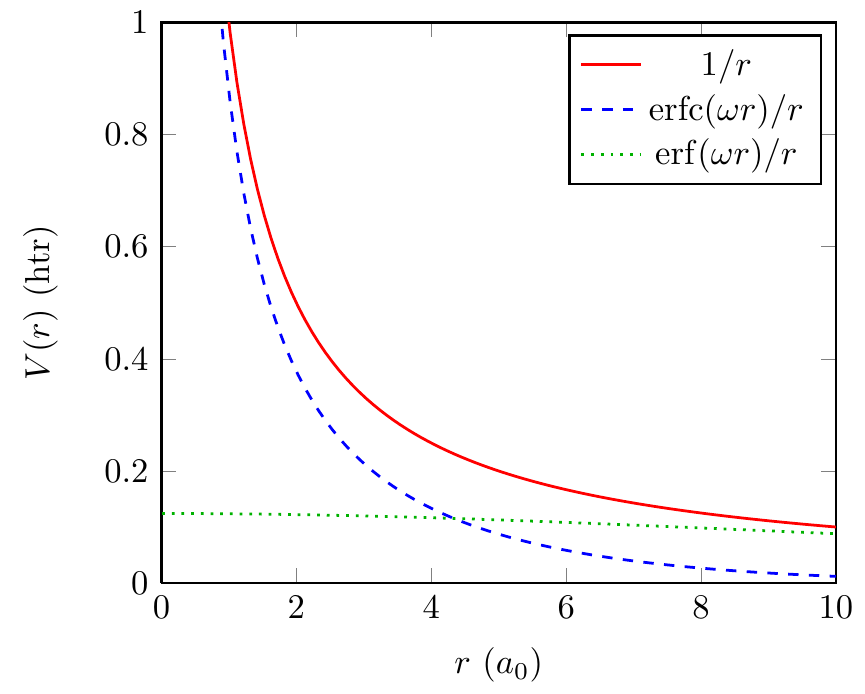}
	\caption{(Color online) Comparison of the bare (red solid line) and the screened (blue dashed line) Coulomb potentials
	 with the difference between both (green dotted line). The difference does not exhibit a divergence at $r=0$ and is
	 a smooth function everywhere. Its Fourier transform converges very rapidly.}
	\label{fig:motivFT}
\end{figure}
It seems that the implementation of the HSE functional is now straightforward: the bare Coulomb potential in
\myeqref{eq:coulMat} is replaced by the screened one and it is proceeded as in Ref.~\onlinecite{bfb10} for the case of the
 PBE0 functional.
However, in this way we would loose a very favorable property of the bare Coulomb potential, namely, its multipole
expansion, which makes it possible to render the Coulomb matrix sparse by a simple unitary transformation of the
MPB. The sparsity of $v_{IJ}(\vect q)$ speeds up the matrix-vector multiplications in \myeqref{eq:introMixed}
considerably. The screened Coulomb potential does not have this simple property. Furthermore, the direct evaluation of
$v_{IJ}(\vect q)$ following the techniques of Ref.~\onlinecite{fsb09} is not transferable to the screened
interaction.
\par
For these reasons, we evaluate \myeqref{eq:introMixed} with the bare interaction as before and use a separate
step to incorporate the screening. This procedure is motivated by \myfigref{fig:motivFT}, which shows the bare and
screened Coulomb potentials, $v(r)$ and $v\apex{SR}(r)$, as well as the difference, $v\apex{LR}(r)$, as a function
of the distance $r$, measured in units of Bohr radii $a_0$. While the two potentials diverge at $r=0$, their difference
 remains
finite. It has a very smooth behavior for all distances and should thus be suitable to be described in reciprocal
space. In fact, we find
that only very few plane waves are needed to represent the difference accurately.\par

To make use of the quickly converging Fourier expansion of the long-range potential $v\apex{LR}(r)$,
 we rewrite
the total xc potential in the form
\begin{equation}
  \potential\pedex{xc}\apex{HSE} =
   \potential\pedex{xc}\apex{PBE} - a \potential\pedex{x}\apex{PBE,SR}
+ a (\potential\pedex{x}\apex{NL} - \potential\pedex{x}\apex{NL,LR} )
\end{equation}
with the local xc potentials
$V\pedex{xc}\apex{PBE}$ and
$V\pedex{x}\apex{PBE,SR}$
that derive from
$E\pedex{xc}\apex{PBE}$ and
$E\pedex{x}\apex{PBE,SR}$,
respectively,
and the potential terms in the parenthesis sum up to
 the SR nonlocal potential for HSE as
\begin{widetext} \begin{equation} \label{eq:sepHSE}
	\potential_{\text{x},nn'}^{\sigma,\text{NL,SR}}(\vect k) = \potential_{\text{x},nn'}^{\sigma,\text{NL}}(\vect k)
+ \sum_{m}\apex{occ} \sum_{\vect q}\apex{BZ}\sum_{\vect G}\braketvec{ \varphi^\sigma_{n\vect k}}{
\varphi^\sigma_{m\vect k - \vect q}\chi_{\vect q + \vect G}}\braketmat{\chi_{\vect q + \vect G}}{ v\apex{LR} }{\chi_{\vect q + \vect
G}}\braketvec{\chi_{\vect q + \vect G} \varphi^\sigma_{m\vect k - \vect q}}{ \varphi^\sigma_{n'\vect k} }\space,
\end{equation} \end{widetext}%
where $\chi_{\vect q + \vect G}(\vect r) = \expp{\imag(\vect q + \vect G) \cdot \vect r} / \sqrt{V}$ is a plane
wave normalized by the crystal volume $V$.
\begin{figure}
  \includegraphics{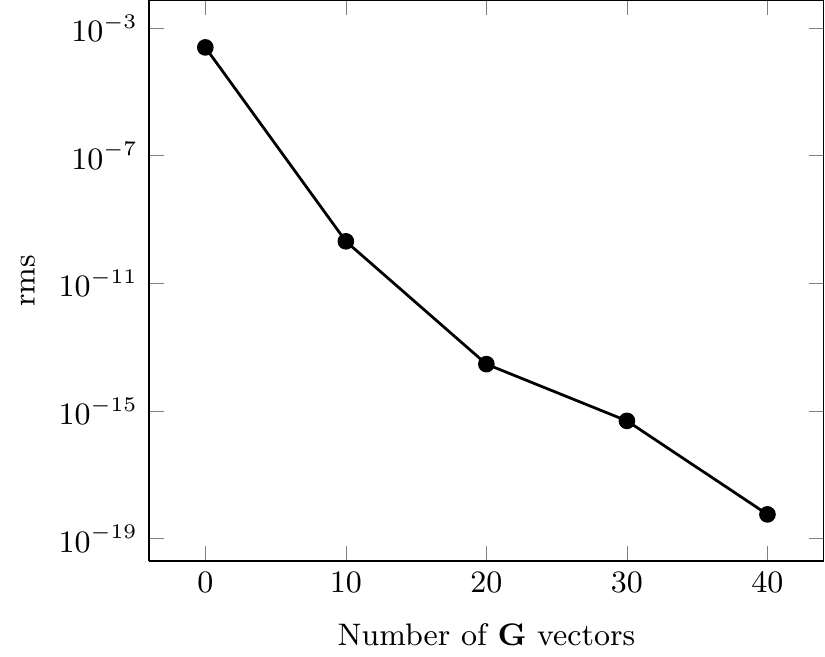}
	\caption{Root-mean-square (rms) deviation of the eigenvalues of the second term in \myeqref{eq:sepHSE} from the fully
		converged result as a function of the number of plane waves in the Fourier transformation for the $\Gamma$ point of
		silicon. We have used a supercell containing eight atoms. The 40 Fourier components would translate to ten in a
		calculation of the primitive unit cell with two atoms.}
	\label{fig:convFT}
\end{figure}%
We evaluate the Fourier transform of the wave-function products with the help of the MPB
\begin{equation} \label{eq:FT_wave}
	\braketvec*{ \varphi^\sigma_{n\vect k}}{\varphi^\sigma_{m\vect k - \vect q}\chi_{\vect q + \vect G}} = \sum_{I} \braketvec*{ \varphi^\sigma_{n\vect k}}{ \varphi^\sigma_{m\vect k - \vect q}\mixed_{\vect qI} } \braketvec*{ \tilde\mixed_{\vect qI} }{\chi_{\vect q + \vect G}}\space,
\end{equation}
where the first integrals on the right-hand side are calculated routinely already for $V^{\sigma,\text{NL}}_{\text{x},nn'}$.
The Fourier transform of $v\apex{LR}$ is known analytically
\begin{equation} \label{eq:FTpotLR}
	\braketmat{\chi_{\vect q + \vect G}}{ v\apex{LR} }{\chi_{\vect q + \vect G'}} = \frac{4\pi}{\abs{\vect q + \vect G}^2} \expp{-\abs{\vect q + \vect G}^2 / 4\omega^2} \delta_{\vect G\vect G'} \space.
\end{equation}
We note that any other form of the screened Coulomb interaction
could easily be implemented at this stage.
Since the matrix elements are diagonal in reciprocal space, the second term
of \myeqref{eq:sepHSE} takes in practical terms negligible time to compute.
From the fact that this function approaches zero very quickly with $\abs{\vect q + \vect G}$, it is clear that the
results are easily converged up to machine precision, even if the Fourier coefficient in \myeqref{eq:FT_wave}
falls off slowly with respect to $\abs{\vect q+\vect G}$
because of the rapidly varying all-electron wave functions. Figure~\ref{fig:convFT} shows the root-mean-square deviation
of the eigenvalues of the matrix
$\potential_{\text{x},nn'}^{\sigma,\text{NL,SR}}(\vect k) -
 \potential_{\text{x},nn'}^{\sigma,\text{NL}}(\vect k)$, as a function of the number of $\vect G$ vectors used for its
 construction. The convergence was studied for the case of bulk silicon using a supercell with eight atoms.
Machine precision is achieved with as few as 40 plane waves which would translate to ten for a primitive unit cell
containing two atoms. This behavior is independent of the $\vect q$ point.\par

We note that the Fourier transform in \myeqref{eq:FTpotLR} diverges as $1/\abs{\vect q + \vect G}^2$ in the limit $\vect q + \vect G \rightarrow \vect 0$.
The same divergence is found for the bare Coulomb potential,\cite{bfb10,fsb09} such that the $1/\abs{\vect q + \vect G}^2$ terms cancel in the difference. The remainder is finite and is given by
\begin{equation}
	\lim_{\vect q + \vect G \rightarrow \vect 0} \frac{4\pi}{\abs{\vect q + \vect G}^2} \left( 1 - \expp{-\abs{\vect q + \vect G}^2 / 4\omega^2} \right) = \frac{\pi}{\omega^2}\space.
\end{equation}
We will later see that this nondivergent behavior of the screened interaction gives rise to a favorable $\vect k$-point convergence.

\section{Verification} \label{sec:calc}
First, we present calculations for a prototypical set of semiconductors (C, Si, and GaAs)
 and insulators (MgO, NaCl, and Ar) and compare the results with previous
works from the literature.\cite{paier06,kvis06} We focus in particular on direct and indirect band transitions. These are
calculated as the energy differences of the highest occupied and the lowest unoccupied eigenstates at
the corresponding points in the BZ. We have taken the experimental lattice
constants from Ref.~\onlinecite{hs04_2}.
\begin{figure}
  \includegraphics{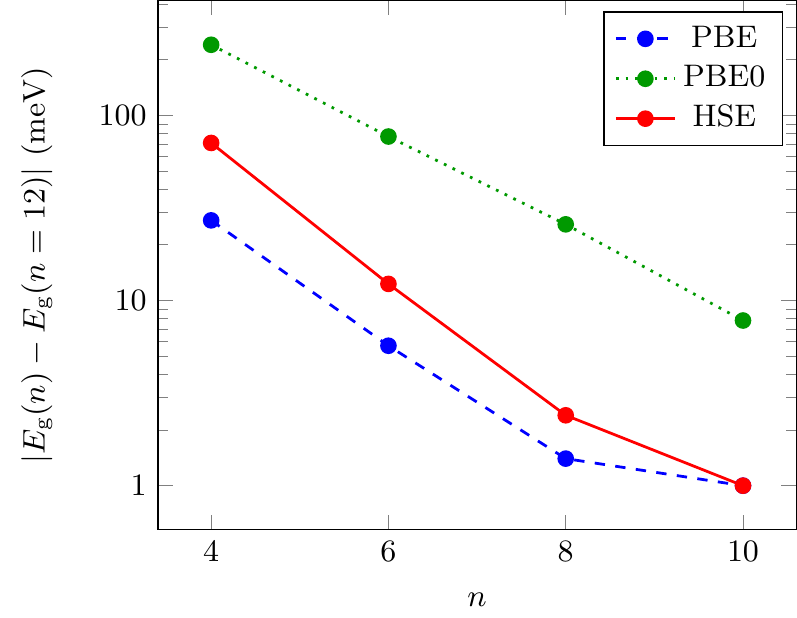}
	\caption{(Color online) Convergence of the silicon band gap $E\pedex{g}$ for the functionals
	PBE (blue, dashed), PBE0 (green, dotted), and HSE (red, solid) with respect to the $\vect k$-point mesh ($n \times n \times n$).}
	\label{fig:kptConv}
\end{figure}
In \myfigref{fig:kptConv}, we show the convergence of the band gap for silicon with the size of the
$\vect{k}$-point mesh. Within HSE this convergence is almost as fast as in PBE, whereas PBE0 requires larger
$\vect{k}$-point meshes. This can be attributed to the nondivergent behavior of the screened
interaction at $\vect k = \vect 0$ (s.~Sec.~\ref{sec:impl}) and was already observed in Ref.~\onlinecite{paier06}.
We find that an 8$\times$8$\times$8 $\vect k$-point mesh gives reliable HSE results
for the band gap as well as for ground-state properties.
\par
\begin{table*}
	\caption{Kohn-Sham transition energies in eV obtained with the functionals PBE and HSE at experimental lattice
constants compared with values from PAW calculations and experiment. An 8$\times$8$\times$8 $\vect k$-point mesh was employed.}
	\begin{ruledtabular}
	\begin{tabular}{l l r r r r c}
		\multicolumn{2}{l}{} & \multicolumn{2}{c}{This work} &
		\multicolumn{2}{c}{PAW\footnote{Reference~\onlinecite{paier06}.}} & Expt. \\
		\multicolumn{2}{l}{} & \multicolumn{1}{c}{PBE} & \multicolumn{1}{c}{HSE} &
		\multicolumn{1}{c}{PBE} & \multicolumn{1}{c}{HSE} \\ \hline
		GaAs & $\Gamma \rightarrow \Gamma$ & 0.54 &  1.43 &  0.56 &  1.45 & 1.52,\footnote{Reference~\onlinecite{hell82}} 1.63\footnote{Reference~\onlinecite{ch89}.}\\
		     & $\Gamma \rightarrow$ X      & 1.47 &  2.06 &  1.46 &  2.02 & 1.90,\footnotemark[2] 2.01,\footnotemark[3] 2.18\footnotemark[3]\\
		     & $\Gamma \rightarrow$ L      & 1.01 &  1.78 &  1.02 &  1.76 & 1.74,\footnotemark[2] 1.84,\footnotemark[3] 1.85\footnotemark[3]\\
		Si   & $\Gamma \rightarrow \Gamma$ & 2.56 &  3.32 &  2.57 &  3.32 & 3.05,\footnote{Reference~\onlinecite{oh93}.} 3.34--3.36,\footnote{Reference~\onlinecite{wb72}.} 3.4\footnotemark[3] \\
		     & $\Gamma \rightarrow$ X      & 0.71 &  1.29 &  0.71 &  1.29 & 1.13,\footnotemark[5] 1.25\footnotemark[4] \\
		     & $\Gamma \rightarrow$ L      & 1.54 &  2.24 &  1.54 &  2.24 & 2.06,\footnote{Reference~\onlinecite{hn76}.} 2.40\footnotemark[3] \\
		C    & $\Gamma \rightarrow \Gamma$ & 5.60 &  6.98 &  5.59 &  6.97 & 7.3\footnotemark[2] \\
		     & $\Gamma \rightarrow$ X      & 4.75 &  5.90 &  4.76 &  5.91 & --- \\
		     & $\Gamma \rightarrow$ L      & 8.46 & 10.02 &  8.46 & 10.02 & --- \\
		MgO  & $\Gamma \rightarrow \Gamma$ & 4.77 &  6.49 &  4.75 &  6.50 & 7.7\footnote{Reference~\onlinecite{ada99}.} \\
		     & $\Gamma \rightarrow$ X      & 9.14 & 10.86 &  9.15 & 10.92 & --- \\
		     & $\Gamma \rightarrow$ L      & 7.93 &  9.69 &  7.91 &  9.64 & --- \\
		NaCl & $\Gamma \rightarrow \Gamma$ & 5.20 &  6.57 &  5.20 &  6.55 & 8.5\footnote{Reference~\onlinecite{pllj75}.} \\
		     & $\Gamma \rightarrow$ X      & 7.58 &  9.05 &  7.60 &  8.95 & --- \\
		     & $\Gamma \rightarrow$ L      & 7.30 &  8.66 &  7.32 &  8.67 & --- \\
		Ar   & $\Gamma \rightarrow \Gamma$ & 8.70 & 10.36 &  8.68 & 10.34 & 14.2\footnote{Reference~\onlinecite{mfg04}.}
	\end{tabular}
	\end{ruledtabular}
	\label{tab:bandgap}
\end{table*}
In \mytabref{tab:bandgap}, we compare our results for the $\Gamma\rightarrow\Gamma$, $\Gamma\rightarrow \text{X}$, and
$\Gamma \rightarrow \text{L}$ transition energies with those obtained by the projector-augmented-wave (PAW)
method\cite{paier06} and experimental data. The band transitions are calculated for a set of materials at their
experimental lattice constants with the functionals PBE and HSE. Overall, we observe excellent agreement between
the calculated values. In comparison to the experimental results, the HSE
functional considerably improves on the PBE values. For semiconductors, the
HSE transition energies are in very good agreement with experiment, while
the larger gaps of insulators are still underestimated.
\par
\begin{table}
	\caption{Optimized lattice constants in \AA~obtained with the PBE and the HSE functional. An
	$8\times8\times8$ $\vect k$-point mesh was employed. Results are compared to experimental results and calculations
	using the HSE functional within a PAW\cite{paier06} and a Gaussian\cite{kvis06} method.}
  \begin{ruledtabular}
	\begin{tabular}{l c c c c c}
		 & \multicolumn{2}{c}{This work} & PAW\footnote{Reference~\onlinecite{paier06}.} & Gaussian\footnote{Reference~\onlinecite{kvis06}.} & Expt.\footnote{Experimental data taken from Ref.~\onlinecite{hs04_2}.}\\
		Functional & PBE & HSE & HSE & HSE & --- \\ \hline
		GaAs & 5.743 & 5.660 & 5.687 & 5.710 & 5.648 \\
		Si   & 5.472 & 5.441 & 5.435 & 5.451 & 5.430 \\
		C    & 3.571 & 3.549 & 3.549 & 3.557 & 3.567 \\
		MgO  & 4.265 & 4.217 & 4.210 & 4.222 & 4.207 \\
		NaCl & 5.703 & 5.627 & 5.659 & 5.645 & 5.595 \\
	\end{tabular}
	\end{ruledtabular}
	\label{tab:latcon}
\end{table}
\begin{table}
	\caption{Bulk moduli in GPa obtained with the PBE and the HSE functional. An
	$8\times8\times8$ $\vect k$-point mesh was employed. Results are compared to experimental results and calculations
	using the HSE functional within a PAW method.\cite{paier06}}
  \begin{ruledtabular}
	\begin{tabular}{l c c c c}
		 & \multicolumn{2}{c}{This work} & PAW\footnote{Reference~\onlinecite{paier06}.} & Expt.\footnote{Experimental data taken from Ref.~\onlinecite{hs04_2}.}\\
		Functional & PBE & HSE & HSE & --- \\ \hline
		GaAs & 64.5 & 79.2 & 70.9 & 75.6 \\
		Si   & 88.9 & 98.0 & 97.7 & 99.2 \\
		C    & 433 & 467 & 467 & 443 \\
		MgO  & 153 & 177 & 169 & 165 \\
		NaCl & 21.3 & 28.8 & 24.5 & 26.6 \\
	\end{tabular}
	\end{ruledtabular}
	\label{tab:moduli}
\end{table}
We compute the equilibrium lattice constants and bulk moduli by calculating
total energies for different lattice constants and fitting the results to the Murnaghan equation of state.\cite{mur44}
In \mytabref{tab:latcon} and \mytabref{tab:moduli}, we compare the lattice constants and bulk moduli
obtained with our implementation of the
HSE functional with results from implementations based on plane-wave (PAW)\cite{paier06} and Gaussian basis
sets.\cite{kvis06} The results of all three methods agree very well.
For example the lattice constants calculated in the FLAPW and PAW methods
differ by less than $3\unit{pm}$.
Except for diamond, the HSE functional yields lattice constants and bulk moduli
in much better agreement with experiment than the semilocal PBE functional, which tends to
overestimate the former and underestimate the latter.

\section{Gadolinium Nitride} \label{sec:gdn}
\subsection{Computational setup}

\begin{table}
  \caption{Numerical parameters used for the calculation of GdN.}
  \begin{ruledtabular}
  \begin{tabular}{lll}
    \multicolumn{1}{c}{parameter} & \multicolumn{2}{c}{value} \\
    \hline
    $\vect k$-point mesh  & $8\times8\times8$ & \\
    muffin-tin radii      & $R\pedex{MT}(\text{Gd}) = 2.33\unit{$a_0$}$ & $R\pedex{MT}(\text{N}) = 1.95\unit{$a_0$}$ \\
    plane-wave cutoffs    & $G\pedex{max} = 4.9\unit{$a_0^{-1}$}$       & $G'\pedex{max} = 3.6\unit{$a_0^{-1}$}$ \\
    angular-momentum      & $l\pedex{max}(\text{Gd}) = 12$              & $l\pedex{max}(\text{N}) = 10$ \\
    \multicolumn{1}{c}{cutoffs} & $L\pedex{max}(\text{Gd}) = 6$         & $L\pedex{max}(\text{N}) = 4$ \\
    local orbitals\footnote{Reference~\onlinecite{bfbg11}.}
                          & \multicolumn{2}{l}{Gd: 5s,5p;7s,7p,6d,5f} \\
                          & \multicolumn{2}{l}{N: 3s,3p,4d,5f}\\
    number of bands & \multicolumn{2}{l}{200}
  \end{tabular}
  \end{ruledtabular}
  \label{tab:inputGdN}
\end{table}

\begin{table*}
	\caption{Comparison of our HSE results for GdN with those from LSDA+$U$ and B3LYP calculations and experiment. The theoretical results
  are given for the optimized lattice constant, unless stated otherwise.}
  \begin{ruledtabular}
	\begin{tabular}{l r r r r r r r}
	& HSE\footnote{At the experimental lattice constant of $4.988\unit{\AA}$.} & HSE
  & LSDA+$U$\footnote{Reference~\onlinecite{duanall}; $U$ optimized for Gd bulk (Ref.~\onlinecite{harm95}).}
  & LSDA+$U$\footnote{Reference~\onlinecite{llcs07}; $U$ chosen to reproduce the experimental direct gap of paramagnetic GdN (Ref.~\onlinecite{hull79}).}
  & LSDA+$U$\footnotemark[1]$^{,}$\footnote{Reference~\onlinecite{tro07}; $U$ chosen to reproduce the experimental direct gap of paramagnetic GdN.}
  & B3LYP\footnote{Insulating solution of Ref.~\onlinecite{doll08}.} & Expt.\\
	\hline
	Lattice constant (\AA) & (4.988) & 4.967 & 4.92 & 5.08 & (4.988) & 5.05 & 4.988\footnote{At room temperature; Ref.~\onlinecite{bc74}.} \\
	Bulk modulus (GPa)     & ---   & 164   & --- & 150 & --- & 159 & 192\footnote{Reference~\onlinecite{lb02}.} \\
	Magnetic moment ($\mu\pedex B$)   & 6.99 & 6.99 & --- & 6.93\footnotemark[1] & --- & 7.0 & 6.88\footnote{Reference~\onlinecite{liss94}}\\
	Direct gap at X (eV) ($T < T\pedex{C}$) & 0.90 & 0.85 & $-$0.16\footnotemark[9] & --- & 0.91 & 1.18\footnote{Extracted from the band structure.} & 0.90\footnote{Reference~\onlinecite{tro07}.} \\
	Direct gap at X (eV) ($T > T\pedex{C}$) & 1.17 & 1.11 & 0.10\footnotemark[9] & 0.98\footnotemark[1] & 1.30 & 1.77\footnotemark[9] & 1.31\footnotemark[10]\\
	Indirect gap $\Gamma$$\rightarrow$X (eV) ($T < T\pedex{C}$) & 0.01 & $-$0.06 & $-$0.45\footnotemark[9] & 0.14\footnotemark[1] & 0.43 & 0.72\footnotemark[9] & --- \\
	Indirect gap $\Gamma$$\rightarrow$X (eV) ($T > T\pedex{C}$) & 0.90 & 0.85 & $-$0.13\footnotemark[9] & 0.69\footnotemark[1] & 0.98 & 1.47\footnotemark[9] & --- \\
  Position of majority $4f$ peak (eV)\footnote{Relative to the top of the valence band.} & $-$6.00 & $-$6.00 & $-$7.8 & $-$8.1\footnotemark[1]$^,$\footnotemark[9] & --- & $-$6.3\footnotemark[9] & $-$7.8\footnote{Reference~\onlinecite{leuen05}.} \\
  Position of minority $4f$ peak (eV)\footnotemark[11] & 6.05 & 6.05 & 6.6 & 5.0\footnotemark[1]$^,$\footnotemark[9] & 4.8\footnotemark[9] & 5.5\footnotemark[9] & 5.5 -- 6.1\footnote{Reference~\onlinecite{yama96} measured for GdX (X = P, As, Sb, and Bi).}
	\end{tabular}
	\end{ruledtabular}
	\label{tab:resGdN}
\end{table*}

GdN crystallizes in the rocksalt structure, with a room-temperature lattice constant of $a\pedex{GdN} =
4.988\unit{\AA}$.\cite{lb02}
The valence band of this material consists of the N $2s$, $2p$ and the Gd $4f$ states. The $4f$ states are only
 half-occupied. The conduction band is
 formed by the Gd $5d$ and $6s$ states as well as the $4f$ states in the minority channel.
\par

We determine the numerical cutoff parameters for the calculations in such a way that the difference between
the total energies calculated at the experimental lattice constant, $a\pedex{GdN}$, and at $1.01 a\pedex{GdN}$
changes by less than $1\unit{meV}$ upon increasing the parameters. In \mytabref{tab:inputGdN}, we list the parameters used for the
GdN unit cell (consisting of two atoms). In particular, we converged the $\vect{k}$-point sampling,
the size of the FLAPW basis, the size of the MPB, and the number of local orbitals.
The latter are additional basis functions that are used to describe semicore states\cite{s91} or to
eliminate the linearization error.\cite{kra97,fsbk06}
\par
In the following, we compare our theoretical HSE results for the
lattice constant, bulk modulus, band gaps, and magnetic moment with previous
calculations and experiment.
In order to compare our band structure results obtained at $0\unit{K}$  with the
experimental results obtained at room-temperature, where GdN is in the paramagnetic state,
we follow the idea that there is no difference between the ferromagnetic and paramagnetic state
for the exchange splitting and the large magnetic moments of the $4f$ electrons. In the paramagnetic
state we rather assume local magnetic $f$ moments that fluctuate in direction with an overall zero
magnetization. Thus, the magnetic polarization of the N states disappears. Dispersive valence and conduction
electrons that exhibit a large group velocity feel at any moment in time a small random potential landscape due
to the exchange potentials of Gd $f$ moments pointing in random directions. Following Ref.~\onlinecite{llcs07}
this can be approximated assuming that in the paramagnetic phase each of  these $s$, $p$, and $d$
valence and conduction states characterized by a $\vect{k}$-point band index are obtained by the averages
of the corresponding spin-up and spin-down energies in the ferromagnetic phase.

 Furthermore, from the total-energy differences we derive the exchange coupling constants
for a Heisenberg spin Hamiltonian and determine the Curie temperature.
This gives us a measure for the quality of energy differences that can be
expected from the HSE functional between different magnetic states.

\subsection{Structural and electronic properties}
\begin{figure*}
  \includegraphics{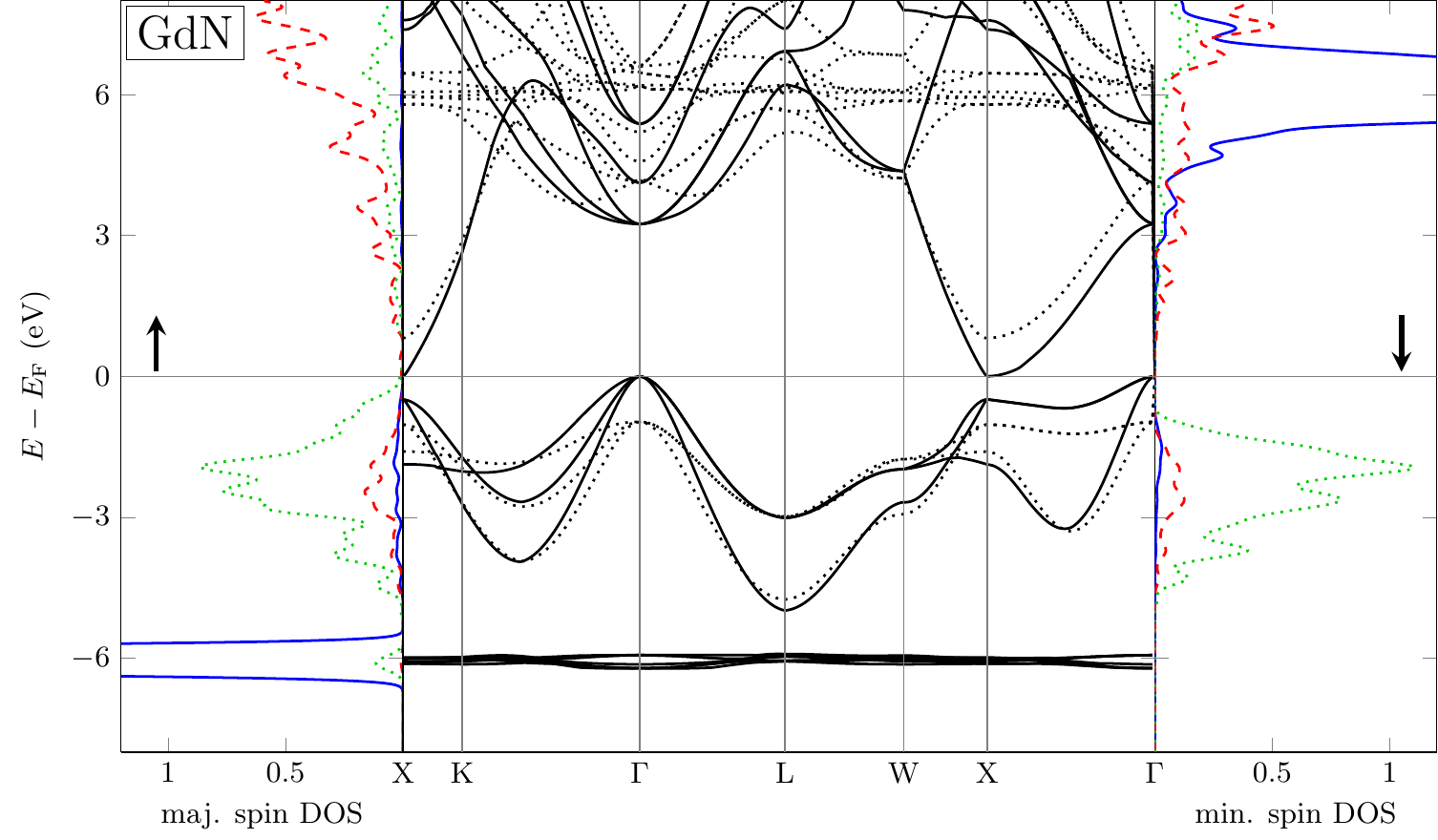}
	\caption{(Color online) Band structure and density of states (DOS; in states per eV) of GdN at the experimental
	lattice constant. The majority and minority bands are plotted as solid and dotted lines, respectively.
  The orbital-resolved DOS is shown on the left for majority and on the right for minority states. The
  solid blue line shows the Gd $4f$ states, the red dashed line the Gd $5d$ states,
	and the green dotted line the N $2p$ states.}
	\label{fig:bandGdN}
\end{figure*}
We start our investigation of GdN by evaluating its structural and electronic properties and comparing them to some
 of the available experimental\cite{liss94,bc74,lb02,tro07} and theoretical data, obtained with the hybrid B3LYP
 functional\cite{doll08} and within the LSDA+$U$ approach.\cite{duanall,llcs07,tro07} The comparison is
shown in \mytabref{tab:resGdN}. We note that our parameter-free HSE calculations yield a lattice parameter
of $4.967\unit{\AA}$  in very close agreement to the experiment, while B3LYP overestimates the
value by $\sim2\%$ and the lattice constant in the LSDA+$U$ method depends on the choice of the parameter $U$.
 Thermal expansion could account for
the remaining difference to the experimental lattice parameter that was determined at the room temperature (whereas the theoretical result corresponds to $0\unit{K}$).
The thermal expansion coefficient of GdN is unknown so far.
Assuming linear expansion between $0\unit{K}$ and room temperature ($293\unit{K}$) with the coefficient of isostructural and isovalent EuO
($\approx 13\times10^{-6} \unit{K}^{-1}$),\cite{levy69} one would extrapolate to
$4.969\unit{\AA}$ at $0\unit{K}$, which is, indeed, very close to our optimized lattice constant.
\par
Next, we turn to the electronic structure. In \myfigref{fig:bandGdN} we show the spin-resolved band structure and the spin- and
orbitally resolved density of states calculated at the experimental lattice
constant ($4.988\unit{\AA}$).
At about $-6\unit{eV}$ and $+6\unit{eV}$ we find the
localized majority and minority Gd 4$f$ state, respectively.
In the vicinity of the Fermi energy, GdN exhibits a truly
interesting electronic structure. Of particular interest are a direct and an indirect band gap, discussed extensively in the literature.
 The direct band gap accessible by optical measurements is located at the X point
 and amounts to $0.9\unit{eV}$ for the majority spin channel and $1.5\unit{eV}$ for the minority spin channel. The valence
and the conduction states are separated by an indirect band gap
($\Gamma$$\rightarrow$$\text{X}$): In the minority-spin states, this is a robust band gap of $1.5\unit{eV}$, while in the majority-spin states this gap is tiny, only $0.01\unit{eV}$. Thus, we
observe that GdN is in a narrow-gap  semiconducting ground state with an almost vanishing indirect band gap
($\Gamma$$\rightarrow$$\text{X}$) in the majority-spin direction, between the N $2p$ states in the valence and the Gd
$5d$ states in the conduction band.
This explains the different experimental reports  disclosing
GdN as a low carrier semimetal or an insulator depending on small changes of the
experimental circumstances.
\par
 Upon decreasing the lattice constant isotropically by just $0.02\unit{\AA}$ to the theoretically optimized value of $4.967\unit{\AA}$,
 we observe a transition to a half-metallic state or, more precisely, to a
semimetallic state just for the majority states: a small portion of the N $2p$ states at the $\Gamma$ point
remains unoccupied, while the Gd $5d$ band becomes partially occupied at the X point. If we
define the ``band gap'' as the difference of these states, we formally get its value to be negative
 (see the band-transition energies in \mytabref{tab:resGdN}).
The described transition from a half-metallic to semiconducting state under isotropic strain was also observed by Duan
 \etal,\cite{duanall} however at a much larger lattice constant of $5.63\unit{\AA}$.
Our results suggest that, since the semiconductor to half-metal transition occurs so close to the equilibrium
lattice parameter, the growth conditions, which influence the material properties such as the concentration of possible N vacancies or strain in the system,
or the lattice expansion upon temperature changes play a decisive role in the transport properties of GdN.
\par
Interesting physics
can be expected also upon doping. If we n-dope into the conduction band or
p-dope into the valence band, e.g. by use of Eu, at a concentration where
electrons or holes populate only majority states, we can obtain significant charge currents
with 100\% spin polarization. In the paramagnetic phase,
minority and majority states converge to spin-degenerate valence and conduction
states, and any spin polarization of the charge current disappears.
At the same time, the band gap opens in the paramagnetic state, partly because of the thermal expansion and partly
because of the averaged-out exchange potential felt by the electrons. This, and a possible coupling of the conduction
electrons or holes to the fluctuating $4f$ moments\cite{ak08} may change
the conductivity by orders of magnitude when passing through the Curie temperature ($T\pedex{C}$).
\par
In \mytabref{tab:resGdN}, we list the band-transition energies calculated with our method at the experimental and
 at the optimized lattice constant for the ferromagnetic ($T<T\pedex{C}$) and the paramagnetic ($T>T\pedex{C}$) state.
 We find invariably larger
band gaps for the paramagnetic state with no transition to a
metallic state nearby. Our calculated band-transition energies compare
 well with the experimental data where available. The small indirect  majority spin gap between
$\Gamma$ and $\text{X}$ of $0.01\unit{eV}$ compares well with estimates of $0.05\unit{eV}$ from Chantis \etal\cite{csk07}
 obtained using the quasiparticle self-consistent $GW$ method (QSGW) combined with their empirical rule
 to estimate band gaps in semiconductors. The transition energies obtained within the LSDA+$U$ method depend
strongly on the choice of the parameter $U$. At the experimental lattice constant, we find similar transition energies as
in the works of Larson \etal\cite{llcs07} and Trodahl \etal,\cite{tro07} where in the latter the parameter $U$
 applied to the $4f$ electrons was chosen to agree with the differences of the binding energies of the occupied and
 unoccupied $4f$ level as measured with the X-ray photoemission and inverse photoemission, respectively, in the
 Gd pnictides and the $U$ applied to Gd $d$ states was chosen to reproduce the
direct experimental band gap in the paramagnetic phase. In this way, the redshift of the direct band gap
of $0.41\unit{eV}$ going from the
paramagnetic state to the ferromagnetic state is perfectly reproduced, while our parameter-free
 calculation gives $0.27\unit{eV}$. The crossing of the conduction and the valence band at the X point
 was obtained
 with smaller values of $U$,\cite{duanall} whereas we find a direct band gap at X even below the Curie temperature
 $T\pedex{C}$. The calculation with B3LYP\cite{doll08} yields three different solutions, where only the
 insulating one is similar to our result. The band gaps are significantly larger which may be attributed to the larger
 optimized lattice constant.
\par
The binding energy of the Gd $4f$ majority band is also improved in the HSE scheme: the partial
 compensation of the self-interaction
error leads to a pronounced shift of the localized $4f$ states to larger binding energies. Calculation with the
PBE functional yields a much too shallow $f$ majority band, located at $3.1\unit{eV}$ below the Fermi energy; in HSE this
band appears at a binding energy of $6.0\unit{eV}$, much closer to its experimentally measured position
at $7.8\unit{eV}$.\cite{leuen05} Furthermore, we note a very good agreement with the insulating B3LYP result, where
the position of the $4f$ peak is found at $6.3\unit{eV}$.\cite{doll08} However, the agreement with experiment
 is not perfect. For the $f$ systems
a stronger mixing of nonlocal exchange would probably give a better result. As compared to PBE results,
 the unoccupied $4f$
 minority states shift towards higher energies ($6.05\unit{eV}$), which is consistent with previous LSDA+$U$
 and B3LYP calculations.\cite{llcs07,duanall,doll08} The position of the unoccupied $4f$ states agrees well
 with typical experimental results for the gadolinium pnictides between $5.5\unit{eV}$ and $6.1\unit{eV}$ obtained with
 inverse photoemission spectroscopy.\cite{yama96}

\subsection{Magnetic order and critical temperature}
The ground state of unconstrained bulk GdN is ferromagnetic (FM), with a Curie temperature
of $58\unit{K}$ \cite{lb02} and a magnetic moment
 of $6.88\unit{$\mu\pedex{B}$}$ per Gd atom\cite{liss94}
 determined from the saturation magnetization at $1.2\unit{K}$.
 The total calculated magnetic moment is $7\unit{$\mu\pedex{B}$}$ per formula unit of which $6.99\unit{$\mu\pedex{B}$}$
comes from the Gd muffin-tin sphere.
Our results are in good agreement with these observations (see the comparison of the theoretical and experimental values of the
 magnetic moment in \mytabref{tab:resGdN}). We confirm the magnetic order and determine the critical temperature by mapping the total energies obtained
 from our HSE calculations for several magnetic structures onto the classical Heisenberg Hamiltonian
\begin{equation}
\hamiltonian = -\frac 12\sum_i \vect S_i \left( J_1 \sum_{j=\text{nn}} \vect S_j + J_2 \sum_{j=\text{nnn}} \vect S_j \right) \space
\end{equation}
including the nearest neighbors (nn), and the next nearest neighbors (nnn) interaction
where $J_1$ and $J_2$ are the respective coupling constants with normalized spin vectors $\vect S_i$ and $\vect S_j$.
Positive and negative values of $J$ favor parallel and antiparallel spin alignment,
respectively.
The coupling constants are extracted from the differences of the total energies of the FM, and two types of
antiferromagnetic (AFM) configurations characterized by  planes of ferromagnetically ordered moments that
are antiferromagnetically stacked along the crystallographic [001] or [111] directions (AFM-I and AFM-II,
 respectively).\cite{duanall}
 For the calculation of the AFM-I (AFM-II)  phase we use a tetragonal $1\times1\times2$ (trigonal
$\sqrt[3]{2}\times\sqrt[3]{2}\times\sqrt[3]{2}$) unit cell containing two Gd atoms and calculate the FM state in the same
unit cells, in order to guarantee reliable total energy differences. All the calculations are performed at
the experimental lattice constant.

\par
From the expressions
\begin{align}
	\Delta E_I &= E\pedex{AFM,$I$} - E\pedex{FM,$I$} = 8 J_1 \label{eq:afm1} \\
	\Delta E_{II} &= E\pedex{AFM,$II$} - E\pedex{FM,$II$} = 6 J_1 + 6 J_2\space, \label{eq:afm2}
\end{align}
the Heisenberg coupling constants $J_1$ and $J_2$ are easily obtained. We list them in \mytabref{tab:compareMagn}, along with
 their values calculated in previous studies using an LSDA+$U$ method within an FLAPW~\cite{duanall} and an LMTO basis
 set.\cite{ml08} Both coupling constants are positive,
 confirming the ferromagnetic nature of the ground state.
 \par
\begin{table}
	\caption{Differences of total energies for different magnetic configurations [Eqs.~\eqref{eq:afm1} and \eqref{eq:afm2}], the Heisenberg coupling constants,
					and the corresponding Curie temperatures within the mean-field approximation [\myeqref{eq:tc_mf}] and random phase
					approximation [\myeqref{eq:tc_rpa}]. Energies and coupling constants are given in meV and the Curie temperatures in K.}
	\begin{ruledtabular}
	\begin{tabular} {l r r r r r r}
	& $\Delta E_I$ & $\Delta E_{II}$ & $J_1$ & $J_2$ & $T\pedex{C}\apex{MFA}$ & $T\pedex{C}\apex{RPA}$ \\
	\hline
	This work & 8.8 & 7.6 & 1.09 & 0.17 & 55 & 42 \\
	Duan \etal\footnote{Reference~\onlinecite{duanall}.} & 6.7 & 4.2 & 0.84 & $-0.14$ & 36 & 26\\
	Mitra \etal\footnote{Reference~\onlinecite{ml08}.} & 3.4 & 0.4 & 0.42 & $-0.36$ & 11 & 5
	\end{tabular}
	\end{ruledtabular}
	\label{tab:compareMagn}
\end{table}

We use two approaches to estimate the Curie temperature. In the mean-field approximation (MFA)
\begin{equation}\label{eq:tc_mf}
	T\pedex{C}\apex{MFA} = \frac{1}{3 k\pedex{B}} \left( 12 J_1 + 6 J_2\right)\space,
\end{equation}
we obtain a $T\pedex{C}\apex{MFA}=55\unit{K}$, very close to the experimental value of $58\unit{K}$.\cite{lb02}
It is known, however, that the mean-field theory overestimates the Curie temperature. For comparison, we have
 also calculated the critical temperature by employing the random phase approximation (RPA) as described in
 Refs.~\onlinecite{bt59} and \onlinecite{now05}, which is known to give results close to Monte Carlo
 solution:
\begin{equation}\label{eq:tc_rpa}
	T\pedex{C}\apex{RPA} = \frac{1}{3 k\pedex{B}} \left[ \int\pedex{BZ} \diff[3]q \frac{1}{J(\vect 0) - J(\vect q)} \right]^{-1}\space,
\end{equation}
where we evaluate the integral on a discrete mesh of $\vect q$ points within the Brillouin zone.
$J(\vect q)$ is the Fourier transform of the exchange coupling constants defined as
\begin{equation}
	J(\vect q) = \sum\pedex{nn} J_1 \expp{\imag\vect q \cdot \vect R\pedex{nn}} + \sum\pedex{nnn} J_2 \expp{\imag\vect q \cdot \vect R\pedex{nnn}}\space,
\end{equation}
where $\vect R\pedex{nn}$ and $\vect R\pedex{nnn}$ are the positions of the nearest and the next nearest neighbors,
respectively. The resulting $T\pedex{C}\apex{RPA} = 42\unit{K}$ is roughly 30\% smaller than the mean-field estimate.
We consider these results as a sophisticated theoretical estimation of the Curie temperature that goes along with
a few uncertainties, some of which are difficult to assess, such as the quality of HSE being an approximation to the true
but unknown exchange and correlation functional and to a much lesser extent the adiabatic approximation inherent in
applying the Heisenberg model. Easier to assess are technically induced error estimates:  (i)~$\Delta E$ in
 Eqs.~\eqref{eq:afm1} and \eqref{eq:afm2} are converged to about $1\unit{meV}$, which translates to an
 uncertainty of
 $3\unit{K}$. (ii)~Based on Monte Carlo calculations with two (as given in \mytabref{tab:compareMagn}) and three
nearest neighbors employing coupling constants published by Duan \etal,\cite{duanall}  which in both cases lead to the same Curie temperature of $28\unit{K}$,
we estimate that the neglect of exchange interactions beyond next nearest neighbors leads to a maximum uncertainty of 1 K.
(iii)~Employing our coupling constants (as given in \mytabref{tab:compareMagn}), we find that the RPA result of
 $42\unit{K}$ approximates the numerically precise determination of the Curie temperature within the
Heisenberg model obtained by Monte Carlo, $45\unit{K}$, by $3\unit{K}$.
\par
 With these error estimates in mind, we compare our values to results of experimental studies, e.g.\ $T\pedex{C}=68$, 69,
 58 or $37\unit{K}$ as reported by Granville \etal,\cite{gra06} Khazen \etal,\cite{khaz06}
 Leuenberger \etal,\cite{leuen05} and Yoshitomi \etal,\cite{yosh11} which vary in value also depending on
 film thickness, strain, grain size, stoichiometry, and N vacancies.\cite{sena11,punya11} We conclude
that our results are in very good agreement with the
experimental situation. Comparing our results to other
 theoretical values exhibited in
 \mytabref{tab:compareMagn} we note that our coupling constants $J_1$ and $J_2$ obtained with HSE are significantly
 higher which gives rise to a higher Curie temperature, in agreement with experiment.
 We observe that the increase of the coupling constants goes along with an increase of the Gd $4f$ moment in the muffin-tin
 sphere by $90\unit{m$\mu\pedex{B}$}$ from $6.78\unit{$\mu\pedex{B}$}$ to $6.87\unit{$\mu\pedex{B}$}$,
a decrease of the Gd $5d$ moment by $20\unit{m$\mu\pedex{B}$}$ from $90\unit{m$\mu\pedex{B}$}$ to $70\unit{m$\mu\pedex{B}$}$,
 and an increase of the N $2p$ moment, which is aligned antiparallel to the Gd $4f$ moment, by $20\unit{m$\mu\pedex{B}$}$,
 from $-100\unit{m$\mu\pedex{B}$}$ to $-120\unit{m$\mu\pedex{B}$}$, when HSE is compared to the PBE functional.
 The precise understanding of the relationship between the change of the moments and the coupling constants requires
 additional analysis that goes beyond the scope of the paper.

\section{Conclusion} \label{sec:conclusion}
In this work, we have presented an implementation of the HSE hybrid functional, which contains a nonlocal screened
exchange potential, within the FLAPW method as implemented in the \texttt{Fleur} code.\cite{fleur} The calculation of
the nonlocal exchange potential is realized by projecting the wave-function products onto the mixed product basis,
reducing the six-dimensional integrations over the nonlocal interaction potential to vector-matrix-vector
products, where the matrix must be calculated only once at the beginning of the self-consistent-field cycle.\par
We employ a sparse-matrix technique\cite{bfb10} to evaluate the vector-matrix-vector products and incorporate the
screening, i.e., the long-range part of the potential, in a separate step, where we exploit its fast converging
Fourier series. This procedure allows constructing the nonlocal HSE potential from PBE0 up to machine precision at a negligible
computational cost. We note that this approach is not restricted to the error function used in the HSE functional.
In fact, our approach is quite general, it can be easily applied to arbitrarily screened interaction potentials.\par
The results for lattice constants and band-transition energies obtained within our method show excellent agreement with
previous results obtained with the PAW\cite{paier06} and Gaussian-based\cite{kvis06}
methods. We have confirmed the finding of Paier
\etal\cite{paier06} that the $\vect k$-point convergence within HSE is comparable to the conventional local PBE
functional, whereas in PBE0 much larger $\vect k$-point meshes are necessary.\par
In addition, we have calculated the properties of the rare-earth compound GdN. There is an ongoing discussion whether the
ground state is insulating or metallic. In fact, within the HSE functional the ground state is very close to
a phase transition: we observe a tiny indirect band gap at the experimental lattice constant at room temperature, which vanishes at the theoretically
optimized $0\unit{K}$ lattice constant -- the compound becomes half-metallic.
The experimentally known band transitions are in good agreement with our theoretical results. Furthermore, we have calculated
the coupling constants for the Heisenberg spin Hamiltonian from total-energy differences of ferromagnetic and
antiferromagnetic configurations. The resulting Curie temperature of $42\unit{K}$ evaluated in the random-phase approximation  is in good
agreement with the experimental value of $58\unit{K}$ and gives confidence
in the energetics obtained by HSE for different magnetic phases. From this we
conclude that the HSE functional has the potential to describe the properties
of rare-earth chalcogenides without the need for employing a
Hubbard $U$ parameter. We encourage
the community to make use of the potential of the HSE functional to explore the
more subtle properties of the rare-earth chalcogenides such as
the physics due to strain, dopands, or heterostructures.

\begin{acknowledgments}
We would like to thank  Walter Lambrecht for fruitful discussions on GdN.
We gratefully acknowledge the funding by the Young Investigators Group Programme of the Helmholtz Association
(``Computational Nanoferronics Laboratory,'' Contract VH-NG-409) and by the Deutsche Forschungsgemeinschaft through
the Priority Program 1145.
\end{acknowledgments}

\bibliographystyle{prsty}
\bibliography{mybib}

\end{document}